\documentclass[pra, twocolumn, showkeys, eqsecnum]{revtex4}
\usepackage{graphics,times}

\newcommand {\psig}{\Sigma_}
\newcommand {\pxi}{\Xi_}
\newcommand {\apsig}[1]{\langle \Sigma_{#1} \rangle}
\newcommand {\apxi}[1]{\langle \Xi_{#1} \rangle}
\newcommand {\apsigxi}[2]{\langle \Sigma_{#1} \Xi_{#2} \rangle}

\newcommand {\Tr} {{\mbox{Tr}}}

\begin{document}
\title{Entanglement increase from local interactions  with
not-completely-positive maps}

\author{Thomas F. Jordan}
\email[email: ]{tjordan@d.umn.edu}
\affiliation{Physics Department, University of Minnesota, Duluth, Minnesota
55812}
\author{Anil Shaji}
\email[email: ]{shaji@unm.edu}
\affiliation{The University of New Mexico, Department of Physics and Astronomy,
800 Yale Blvd. NE, Albuquerque, New Mexico 87131}
\author{E. C. G. Sudarshan}
\email[email: ]{sudarshan@physics.utexas.edu}
\affiliation{The University of Texas at Austin, Center for Statistical
Mechanics, 1 University Station C1609, Austin Texas 78712}  

\begin{abstract}
Simple examples are constructed that show the entanglement of two qubits being both increased and decreased by interactions on just one of them. One of the two qubits interacts with a third qubit, a control, that is never entangled or correlated with either of the two entangled qubits and is never entangled, but becomes correlated, with the system of those two qubits. The two entangled qubits do not interact, but their state can change from maximally entangled to separable or from separable to maximally entangled. Similar changes for the two qubits are made with a swap operation between one of the qubits and a control; then there are compensating changes of entanglement that involve the control. When the entanglement increases, the map that describes the change of the state of the two entangled qubits is not completely positive. Combination of two independent interactions that individually give exponential decay of the entanglement can cause the entanglement to not decay exponentially but, instead, go to zero at a finite time. 
\end{abstract}



\maketitle

\section{Introduction}\label{one}

We construct simple examples here that show the entanglement of two qubits being
both increased and decreased by interactions on just one of them. In our first
and basic step, taken in Sec.~II, we have one of the two qubits interact with
a third qubit, a control, that is never entangled or correlated with either of
the two entangled qubits and is never entangled, but becomes correlated, with
the system of those two qubits. In Sec.~III, we do this for each of the two
entangled qubits, and consider the combination of the two interactions, with
separate control qubits that are not correlated and do not interact with each
other. The two entangled qubits do not interact, but their state can change from
maximally entangled to separable or from separable to maximally entangled.
Similar changes for the two qubits are made with a swap operation between one of the qubits and a control; then there are compensating changes of entanglement that involve the control. This is described in Sec.~II.A.

Whenever the entanglement increases, and in some cases where the entanglement
decreases, the map that describes the change of the state of the two entangled
qubits is not completely positive and does not apply to all states of two
qubits. It all depends on whether there are correlations with the controls at
the beginning of the interval for which the dynamics is considered. The maps are
described in Sec.~IV and discussed in Sec.~V. The completely positive maps
that decrease the entanglement have already been described  \cite{ziman05a}.

When the interaction of each qubit with its control by itself gives exponential
decay of the entanglement, the combination of the two interactions gives
exponential decay at the rate that is the sum of the rates for the individual
interactions, when the two interactions are made the same way. Making them
differently can cause the entanglement to not decay at that rate or at any
single rate. Instead, the entanglement goes to zero at a finite time; the state
becomes separable and remains separable at later times. This is described in
Sec.~III.A. Similar behavior has been observed in more physically interesting
and mathematically complicated models \cite{yu06a,liang06a,yu07a}.

These examples are built on the same framework, but to a very different design,
from those we made for Lorentz transformations that entangle spins
\cite{jordan06b}. There the momenta that played the roles of controls were
purposely correlated. Here the controls are kept independent. The framework
makes the operations transparent by describing the qubit states with density
matrices written in terms of Pauli matrices, so you can see the Pauli matrices
being rotated by the interactions. States are shown to be separable by writing
out the density matrices explicitly as sums of products for pure states. For
each interaction here, the map that makes the change of the density matrix for
the entangled qubits is described by a simple rule that particular Pauli
matrices in the density matrix are multiplied by a number; equivalently, the map
of the state of the entangled qubits is described by a rule that particular mean
values are multiplied by a number.

Our examples show that statements like ``entanglement should not increase under
local operations and classical communication" \cite{bennett96c,horodecki07a}
are not generally true outside the set of local operations considered in
the original proof  \cite{bennett96c}. In our examples, each control qubit
interacts with only one of the two entangled qubits. In this sense, the quantum
operations are local. Correlation with a control at the beginning of the
interval for which the dynamics is considered can give local operations that
increase entanglement.

\section{One interaction}\label{two}  

We consider the entanglement of two qubits, $A$ and $B$. We use Pauli matrices
$\Sigma_1 , \Sigma_2 , \Sigma_3 $ for qubit $A$, and Pauli matrices $\Xi_1 ,
\Xi_2 , \Xi_3 $ for qubit $B$. We let qubit $A$ interact with a third qubit,
which we call $C$. We think of $C$ as a control. By interacting with qubit $A$,
it will control the entanglement of qubits $A$ and $B$. We work with states
represented by orthonormal vectors $|\alpha \rangle $ and $|\beta  \rangle $ for
$C$. We consider a state of the three qubits represented by a density matrix
\begin{equation}
\label{rabdm}
\Pi  = \rho \otimes \frac{1}{2}\openone_C  
\end{equation}
with $\rho $ the density matrix for the state of qubits $A$ and $B$.

We follow common physics practice and write a product of operators for separate systems, for example a product of Pauli matrices $\Sigma$ and $\Xi$ for qubits $A$ and $B$, simply as $\Sigma \Xi$, not $\Sigma \otimes \Xi$. Occasionally we insert a $\otimes$ for emphasis or clarity. We write $\openone_A$, $\openone_B$, $\openone_C$, but we do not put labels $A$ and $B$ on the $\Sigma_j$ and $\Xi_k$. The single statement that the $\Sigma_j$ are for qubit $A$ and the $\Xi_k$ are for qubit $B$ eliminates the need for continual use of both $A$ and $B$ lalels and $\otimes $ signs. 
 
Suppose $\rho $ is one of the density matrices
\begin{equation}
\label{rhopm}
\rho _\pm   = \frac{1}{4}(\openone \pm  \Sigma _1\Xi _1 \pm  \Sigma _2\Xi _2 -
\Sigma _3\Xi _3).
\end{equation}
Both $\rho _+ $ and $\rho _- $ represent maximally entangled pure states for the
two qubits. They are Bell states. The state of zero total spin is represented by
$\rho _- $ and the state obtained from that by rotating one of the spins by $\pi
$ around the $z$ axis is represented by $\rho _+ $. 

For a rotation $W$, let $D_A (W)$ be the $2\times 2$ unitary rotation matrix
made from the $\Sigma_j $ so that
\begin{equation}
D_A (W)^\dagger \, \mathbf\Sigma D_A (W) = W(\mathbf{\Sigma })
\end{equation}
where $W(\mathbf{\Sigma })$ is simply the vector $\mathbf{\Sigma }$ rotated by
$W$. Let $W(\phi   )$ be the rotation by $\phi   $ around the $z$ axis, and let
$D_A (\phi   )$ be $D_A (W(\phi   ))$. 

We consider an interaction between qubits  $A$ and $C$ described by the unitary
transformation
\begin{equation}
\label{UDA}
U = D_A (\phi )|\alpha \rangle \langle \alpha |  + 
D_A (-\phi )|\beta \rangle \langle \beta | 
\end{equation}
or, in Hamiltonian form,
\begin{equation}
\label{UHt}
U = e^{-i\phi H} 
\end{equation}
with
\begin{equation}
\label{Ham}
H = \Sigma_3  \frac{1}{2}(|\alpha \rangle \langle \alpha |  - 
|\beta \rangle \langle \beta |). 
\end{equation}
This changes the density matrix $\rho $ for qubits $A$ and $B$ to
\begin{eqnarray}
\label{UD}
\rho' & = & \Tr_C \big[(U\otimes \openone_B )\Pi (U\otimes \openone_B )^\dagger
\big] \nonumber \\ 
& = & \frac{1}{2}D_A(\phi )\rho D_A(\phi )^\dagger  + \frac{1}{2}D_A(-\phi )\rho
D_A(-\phi )^\dagger. \qquad 
\end{eqnarray}
For $\rho _\pm $ this gives
\begin{eqnarray}
\label{LTr}
\rho _\pm '  & = & \bigg(\frac{1}{2}\bigg)\frac{1}{4}[\openone  \pm   (\Sigma _1
\cos\phi +\Sigma _2 \sin\phi )\Xi _1  \nonumber \\ 
& &  \hspace{1.5 cm} \pm \: (-\Sigma _1 \sin\phi+\Sigma _2
\cos\phi)\Xi _2  - \Sigma _3\Xi _3] \nonumber \\
 & + & \bigg(\frac{1}{2}\bigg)\frac{1}{4}[\openone  \pm  (\Sigma _1
\cos\phi-\Sigma _2 \sin\phi)\Xi _1  \nonumber \\ 
& &  \hspace{1.5 cm} \pm   \: (\Sigma _1 \sin\phi+\Sigma _2 \cos\phi)\Xi _2  -
\Sigma _3\Xi _3] \nonumber \\ 
 & = & \frac{1}{4} [\openone \pm (\Sigma_1 \Xi_1 + \Sigma_2 \Xi_2 ) \cos \phi  -
\Sigma_3 \Xi_3 ] \nonumber \\
 & = & \rho _\pm \cos^2(\phi /2)\; + \; \rho _\mp \sin^2(\phi /2).
\end{eqnarray}

\subsection{From maximally entangled to separable and back}\label{AA}

We focus first on the case where $\phi $ is $\pi /2$. Then both $\rho _+ $ and
$\rho _- $ are changed to 
\begin{eqnarray}
\label{LTrsimp}
\rho ' & = & \frac{1}{4}[\openone - \Sigma _3\Xi _3] \nonumber \\
 & = & \bigg(\frac{1}{2} \bigg)\frac{1}{2}(\openone-\Sigma
_3)\frac{1}{2}(\openone+\Xi_3) \nonumber \\
&& \hspace{1 cm}+\bigg(\frac{1}{2}\bigg)
\frac{1}{2}(\openone+\Sigma _3)\frac{1}{2}(\openone-\Xi _3).
\end{eqnarray}
The density matrix $\rho $ for a maximally entangled state is changed to the
density matrix $\rho ' $ for a separable state that is a mixture of just two
products of pure states. The inverse of the unitary dynamics of qubits $A$ and
$R$ takes $\rho ' $ back to $\rho $; it changes a separable state to a maximally
entangled state.

The dynamics continuing forward also changes this separable state to a maximally
entangled state. As $\phi $ goes from $\pi /2$ to $\pi $, the density matrix
$\rho _\pm '$ changes from that of Eq.~(\ref{LTrsimp}) to
\begin{equation}
\rho _\pm ' = \rho _\mp . 
\end{equation}
There can be revivals of entanglement between two qubits when there is no
interaction between them, as well as when \cite{zyczkowski01a} there is.

Changes in the state of qubits $A$ and $B$ from maximally entangled to
separable and back to maximally entangled can also be made very simply with a
swap of states\cite{zukowski93a} between $A$ and $C$. This can be done with a
unitary operator $U\otimes \openone_B$ with $U$ a unitary operator for qubits
$A$ and $C$ that acts on a basis of product state vectors simply by
interchanging the states of $A$ and $C$. There is interaction between qubits $A$
and $C$ only; qubit $B$ is not involved.

Applied to an initial state described by Eqs.~(\ref{rabdm}) and (\ref{rhopm}),
where qubits $A$ and $B$ are maximally entangled, this swap operation gives a
separable state for $A$ and $B$. Applied a second time, it restores the initial
state where $A$ and $B$ are maximally entangled. For qubits $A$ and $B$, this is
similar to what happens when $\phi $ goes from $0$ to $\pi /2$ to $\pi $. For
the three qubits, it is different. The swap operation does not change the
complete inventory of entanglements for the three qubits. It just moves the
entanglements around. In particular, $C$ becomes maximally entangled with $B$.
We will see, in Secs.~II.C and D, that the interaction described by
Eqs.~(\ref{UDA}), (\ref{UHt}) and (\ref{Ham}) does change the complete inventory
of entanglements for the three qubits. When the state of qubits $A$ and $B$
changes from maximally entangled to separable and back to maximally entangled,
there are no compensating changes of other two-part entanglements. In
particular, qubit $C$ never becomes entangled with anything.

\subsection{Concurrence}\label{B}

The change of entanglement is smaller when $\phi $ does not change by $\pi /2$.
>From Eq.~(\ref{LTr}) we have
\begin{equation}
\label{LTr12}
\rho _\pm '  = \frac{1}{4} [\openone \pm (\Sigma_1 \Xi_1 + \Sigma_2 \Xi_2 ) \cos
\phi + (\Sigma_1 \Xi_1 )(\Sigma_2 \Xi_2 ) ], 
\end{equation}
after rewriting the last term. This shows that for both $\rho_+ ' $ and  $\rho_-
' $ the eigenvalues are
\begin{equation}
\label{eigenval}
\frac{1}{2} (1+\cos\phi ), \: \frac{1}{2} (1-\cos\phi ), \: 0, \: 0 
\end{equation}
because $\Sigma_1 \Xi_1 $ and $\Sigma_2 \Xi_2 $ each have eigenvalues $1$ and
$-1$ and together they make a complete set of commuting operators: their four
different pairs of eigenvalues label a basis of eigenvectors for the space of
states for the two qubits. The Wooters concurrence  \cite{wootters98a} is a
measure of the entanglement in a state of two qubits. It is defined by
\begin{equation}
\label{eq:note10}
C(\rho ) \equiv {\mbox{max}} \left[ 0, \: \sqrt{\lambda_1} -
\sqrt{\lambda_2}-\sqrt{\lambda_3} - \sqrt{\lambda_4} \right]
\end{equation}
where $\rho $ is the density matrix that represents the state and $\lambda_1 ,
\lambda_2 , \lambda_3 , \lambda_4 $ are the eigenvalues, in decreasing order,
of $\rho  \: \Sigma_2 \Xi_2 \: \rho^\star  \:  \Sigma_2 \Xi_2 $, with
$\rho^\star $ the complex conjugate that is obtained by changing $\Sigma_2 $ and
$\Xi_2 $ to $-\Sigma_2 $ and $-\Xi_2 $. From Eq.~(\ref{LTr12}) we have
\begin{equation}
\rho_\pm '  \, \Sigma_2 \Xi_2 \, (\rho_\pm ' )^\star  \, \Sigma_2
\Xi_2 = \rho_\pm ' \, (\rho_\pm ')^\star \, (\Sigma_2 \Xi_2)^2  \!= \!(\rho_\pm
')^2 
\end{equation}
so for $\rho_\pm ' $ the $\sqrt{\lambda_i} $ are the eigenvalues of $\rho_\pm '
$ and the concurrence is
\begin{equation}
\label{concur}
C(\rho_\pm ' ) = |\cos\phi |.
\end{equation}

We can consider the change of entanglement as $\phi $ changes through any
interval. When $|\cos\phi |$ decreases, the entanglement decreases. When
$|\cos\phi |$ increases, the entanglement increases.

\subsection{Two-part entanglements}\label{D}

The only two-part entanglements are when qubit $A$ is in one part and qubit $B$
is in the other. There is entanglement between qubit $A$ and the subsystem of
two qubits $B$ and $C$ and between qubit $B$ and the subsystem of two qubits $A$
and $C$, as well as between qubits $A$ and $B$. There is never entanglement
between the state of qubit $C$ and the state of the subsystem of two qubits $A$
and $B$. The density matrix
\begin{eqnarray}
\label{fulltran}
(U\otimes \openone_B )\Pi (U\otimes \openone_B )^\dagger \!\!\!& =
& \!\!\!\frac{1}{2}D_A(\phi )\rho_\pm  D_A(\phi )^\dagger |\alpha \rangle
\langle \alpha | \nonumber \\
&& \! + \frac{1}{2}D_A(-\phi
)\rho_\pm  D_A(-\phi )^\dagger |\beta \rangle \langle \beta | \nonumber \\
\end{eqnarray}
is always a mixture of two products of pure states. The reduced density matrix
for the subsystem of qubits $A$ and $C$, obtained by taking the trace over the
states of qubit $B$, is just $\openone_A \otimes \openone_C/4 $, and the
reduced density matrix for qubits $B$ and $C$, obtained by taking the trace over
the states of qubit $A$, is $\openone_B \otimes \openone_C/4 $. There is never
entanglement or correlation between qubits $A$ and $C$ or between qubits $B$ and
$C$. The reduced density matrices for the individual single qubits are just
$\openone_A/2 $, $\openone_B/2 $, and $\openone_C/2 $. The only subsystem
density matrix that carries any information is the density matrix $\rho $ for
the qubits $A$ and $B$, which is changed by the interaction with qubit $C$. The
entropy of the subsystem of qubits $A$ and $B$ can increase or decrease, but
there is no change of entropy for any other subsystem or for the entire system
of three qubits.

\subsection{Three-part entanglement}\label{E}

There is three-part entanglement. The state represented by the density matrix
(\ref{fulltran}) is called biseparable because it is separable as the state of a
system of two parts, with $C$ one part and the subsystem of two qubits $A$ and
$B$ the other part. It is not separable as the state of a system of three parts
$A$, $B$, and $C$. The density matrix (\ref{fulltran}) is not a mixture of
products of density matrices for pure states of the individual qubits $A$, $B$,
and $C$. If it were, its partial trace over the states of $C$, the reduced
density matrix that represents the state of the subsystem of the two qubits $A$
and $B$, would be a mixture of products for pure states of $A$ and $B$. That
happens only when $\cos\phi $ is $0$. In that case, we can see that the density
matrix (\ref{fulltran}) is not a mixture of products for pure states of the
individual qubits $A$, $B$, and $C$ because its partial transpose obtained by
changing $\Xi_2 $ to $-\Xi_2 $ is not a positive matrix.

In the classification of three-part entanglement for qubits, biseparable states
are between separable states and states that involve $W$ or $GHZ$ entanglement
 \cite{dur99a,dur00a,acin01a}. Let
$\Pi_1 $, $\Pi_2 $, $\Pi_3 $ be Pauli matrices for the qubit $C$ such that
$|\alpha \rangle \langle \alpha |$ is $(1/2)(\openone+\Pi_3)$ and $|\beta
\rangle \langle \beta |$ is $(1/2)(\openone-\Pi_3)$.  Bounds from Mermin witness
operators say that for separable or biseparable states
\begin{equation}
\label{mermin}
-2 \leq  \langle  \Sigma_j \Xi_j \Pi_j \! -\! \Sigma_j \Xi_k \Pi_k \!  - \!
\Sigma_k \Xi_j \Pi_k \! - \! \Sigma_k \Xi_k \Pi_j \rangle  \leq  2
\end{equation}
for $j,k = 1,2,3$ and $j\not= k$; a mean value outside these bounds is a mark of
$W$ or $GHZ$ entanglement  \cite{toth05a}. In our examples,
these mean values are always $0$. A mean value $\langle |GHZ\rangle \langle
GHZ|\rangle $ larger than $3/4$ for the projection operator onto the $GHZ$
state,
\begin{equation}
\label{GHZ}
|GHZ\rangle  = \frac{1}{\sqrt{2}}|0\rangle |0\rangle |0\rangle +
\frac{1}{\sqrt{2}}|1\rangle |1\rangle |1\rangle , 
\end{equation}
is a mark of $GHZ$ entanglement; it can not be larger than $3/4$ for a $W$ state
 \cite{acin01a}. A mean value $\langle |GHZ\rangle \langle
GHZ|\rangle $ larger than $1/2$ is a mark of a $W$ state; it can not be larger
than $1/2$ for a biseparable state  \cite{acin01a}. In our
examples, $\langle |GHZ\rangle \langle GHZ|\rangle $ is always $0$. A mean value
$\langle |W\rangle \langle W|\rangle $ larger than $2/3$ for the projection
operator onto the $W$ state,
\begin{equation}
\label{W}
|W\rangle  = \frac{1}{\sqrt{3}}|1\rangle |0\rangle |0\rangle +
\frac{1}{\sqrt{3}}|0\rangle |1\rangle |0\rangle  + \frac{1}{\sqrt{3}}|0\rangle
|0\rangle |1\rangle , 
\end{equation}
is a mark of $W$ entanglement; it can not be larger than $2/3$ for a biseparable
state   \cite{acin01a}. In our examples, 
\begin{equation}
\langle |W\rangle \langle W|\rangle  = \frac{1}{6} (1 \pm  \cos\phi ).
\end{equation}
This mean value does not involve either entanglement or correlation of the qubit
$C$; it would be the same if both $|\alpha \rangle \langle \alpha |$ and $|\beta
 \rangle \langle \beta |$ in the density matrix (\ref{fulltran}) were replaced
by $(1/2)_C $, the completely mixed density matrix for $C$.

For any $\phi $, the density matrices (\ref{fulltran}) for the two cases $+$ and
$-$ are changed into each other by the local unitary transformation that changes
the Pauli matrices for one of the qubits $A$ or $B$ by rotating its spin by $\pi
$ around the $z$ axis. As a function of $\phi $, the mean value $\langle
|W\rangle \langle W|\rangle $ changes in opposite directions for the $+$ and $-$
cases. So will any mean value for the states described by the density matrices
(\ref{fulltran}), if it changes at all.

For the states described by the density matrices (\ref{fulltran}), the only
nonzero mean values that involve the qubit $C$ are
\begin{eqnarray}
\label{nz3}
\langle \Sigma _1\Xi _2\Pi_3 \rangle & = & \mp \sin\phi  \nonumber \\
\langle \Sigma _2\Xi _1\Pi_3 \rangle & = & \pm \sin\phi.
\end{eqnarray}
These would be the same if they were calculated with only the $|\alpha \rangle
\langle \alpha |$ part or only the $|\beta  \rangle \langle \beta |$ part of the
density matrix (\ref{fulltran}). In fact, they are the same as $\langle \Sigma
_1\Xi _2 \rangle \langle \Pi_3 \rangle $ and $\langle \Sigma _2\Xi _1 \rangle
\langle \Pi_3 \rangle $ calculated for one of those parts. Their values do not
require either entanglement or correlation of $C$.

\section{Two Interactions}\label{three}

If a control were coupled similarly to qubit $B$ as well, then $\cos\phi $ would
be replaced by $\cos\phi_A \, \cos\phi_B $ in the next to last line of
Eq.~(\ref{LTr}) and in Eqs.(\ref{LTr12}) and (\ref{concur}). If the coupling of
qubit $B$ is made with a rotation around the $x$ axis instead of the $z$ axis,
then the next to last line of Eq.~(\ref{LTr}) becomes
\begin{eqnarray}
\label{LTr2}
\rho _\pm '  \!&=& \! \frac{1}{4} [\openone \pm \Sigma_1 \Xi_1 \cos\phi_A \pm 
\Sigma_2 \Xi_2 \cos \phi_A \cos\phi_B \nonumber \\
&& \hspace{3.5 cm}  - \Sigma_3 \Xi_3
\cos\phi_B].
\end{eqnarray}
Rewriting the last term and looking at eigenvalues in terms of $\Sigma_1 \Xi_1 $
and $\Sigma_2 \Xi_2 $ as before yields the concurrence
\begin{equation}
\label{concur2}
C(\rho_\pm ' )\!\! =\!\! \frac{1}{2} \max[0, \, |\cos\phi_A | + |\cos\phi_A
\cos\phi_B |
+ |\cos\phi_B |  - 1].
\end{equation}
When $\cos\phi_A $ is $1$, these Eqs.(\ref{LTr2}) and (\ref{concur2}) describe
the result obtained when there is only the interaction of qubit $B$ made with a
rotation around the $x$ axis. If neither $\cos\phi_A $ nor $\cos\phi_B $ is $1$,
the concurrence becomes zero, the state separable, before $\cos\phi_A $ or
$\cos\phi_B $ is zero. The interactions of qubits $A$ and $B$ with their
controls change maximally entangled states to separable states. The inverses
change separable states to maximally entangled states. In the following
subsection, we describe the density matrices that show explicitly that the
separable states are mixtures of products of pure states.

\subsection{Exponential decay}\label{A}

To describe exponential decay of entanglement we let
\begin{equation}
\label{gammas}
\cos\phi_A  = e^{-\Gamma_A t}, \; \; \cos\phi_B  = e^{-\Gamma_B t}
\end{equation}
by letting each interaction be modulated by a time-dependent Hamiltonian $H(t)$
that is related to the Hamiltonian $H$ of Eqs.(\ref{UDA}) and (\ref{UHt}) by
\begin{equation}
\label{HtoH}
H(t) = H\frac{d\phi  }{dt} = H\Gamma cot\phi ,  
\end{equation}
where $\phi $ and $\Gamma $ are $\phi_A $ and $\Gamma_A $ or $\phi_B $ and
$\Gamma_B $. The same result could be produced in different ways. The
interactions could be with large reservoirs instead of qubit controls
 \cite{yu06a,liang06a,yu07a}. Each qubit $A$ or $B$ could interact with a stream
of reservoir
qubits  \cite{ecgs03b}. Here we are interested in the way the entanglement
is changed by the combination of the two interactions. That depends only on the
changes in the density matrix $\rho $ for qubits $A$ and $B$, not on the nature
of the controls and the interactions. Maps that make the changes in $\rho $ will
be described in the next section.

If there is only the interaction of qubit $A$ with qubit $C$, the concurrence is
$e^{-\Gamma_A t} $. If there is only interaction of qubit $B$ with its control,
the concurrence is $e^{-\Gamma_B t} $. If there are both and both are made with
rotations around the $z$ axis, the concurrence is $e^{-\Gamma_A t} e^{-\Gamma_B
t} $. If there are both and the interaction of qubit $B$ with its control is
made with a rotation around the $x$ axis, the concurrence is
\begin{equation}
\label{concurG}
C(\rho_\pm ' ) \!= \!\frac{1}{2} \max[0, \; e^{-\Gamma_A t} + e^{-\Gamma_A
t}e^{-\Gamma_B t} + e^{-\Gamma_B t}  - 1].
\end{equation}
This concurrence (\ref{concurG}) is zero when
\begin{equation}
\label{concurzero}
e^{-\Gamma_A t} + e^{-\Gamma_A t}e^{-\Gamma_B t} + e^{-\Gamma_B t}  = 1.
\end{equation}
Then the state is separable; it is a mixture of six products of pure states:
from Eqs.(\ref{LTr2}) and (\ref{gammas}) 
\begin{eqnarray}
\label{mixprod}
\rho_\pm ' & = & \frac{1}{2}e^{-\Gamma_A t}\frac{1}{2}(\openone+\Sigma
_1)\frac{1}{2}(\openone \pm \Xi _1) \nonumber \\
&&  + \frac{1}{2}e^{-\Gamma_A t}\frac{1}{2}(\openone -\Sigma
_1)\frac{1}{2}(\openone \mp \Xi _1) \nonumber \\
 && +  \frac{1}{2}e^{-\Gamma_A t}e^{-\Gamma_B t}\frac{1}{2}(\openone +\Sigma
_2)\frac{1}{2}(\openone \pm \Xi _2) \nonumber \\
&& + \frac{1}{2}e^{-\Gamma_A t}e^{-\Gamma_B t}\frac{1}{2}(\openone -\Sigma
_2)\frac{1}{2}(\openone \mp \Xi _2) \nonumber \\
& & + \frac{1}{2}e^{-\Gamma_B t}\frac{1}{2}(\openone +\Sigma
_3)\frac{1}{2}(\openone -\Xi _3) \nonumber \\
&& + \frac{1}{2}e^{-\Gamma_B t}\frac{1}{2}(\openone -\Sigma
_3)\frac{1}{2}(\openone +\Xi _3).
\end{eqnarray}
The state remains separable at later times; when the sum of the exponential
decay factors is less than $1$, the density matrix is a mixture in which just a
multiple of the density matrix $1/4$ for the completely mixed state is added to
the terms of Eq.~(\ref{mixprod}). This change of maximally entangled states to
separable states can be described without reference to exponential decay by
continuing to use $\cos\phi_A $ and $\cos\phi_B $ instead of $e^{-\Gamma_A t}$
and $e^{-\Gamma_B t}$. Similar behavior involving exponential decay has been
observed in more physically interesting and mathematically complicated models
 \cite{yu06a,liang06a,yu07a}.

\section{Maps}\label{four}

The maps that make the changes in the density matrix $\rho $ for qubits $A$ and
$B$ could be described in different ways using various matrix forms. That is
not needed here. Writing $\rho $ in terms of Pauli matrices provides a very
simple way to describe the maps. For any density matrix
\begin{equation}
  \label{eq:tq1}
  \rho  = \frac{1}{4} \bigg( \openone + \sum_{j=1}^3 \apsig{j} \psig j +
\sum_{k=1}^3 \apxi{k} \pxi k + \sum_{j,k=1}^3 \apsigxi{j}{k} \psig j \pxi k
\bigg)
\end{equation}
for qubits $A$ and $B$, the result of the interaction of qubit $A$ with
qubit $C$, described by Eq.~(\ref{UD}), is that in $\rho $, in both the $\Sigma_j $ and $\Sigma_j \Xi _k$ terms,
\begin{equation}
\Sigma_1 \longrightarrow \Sigma_1 \cos\phi_A , \: \: \Sigma_2 \longrightarrow  \Sigma_2 \cos\phi_A ;
\end{equation}
the result of the interaction of qubit $B$ with
its control is that in $\rho $
\begin{equation}
\Xi _1 \longrightarrow \Xi _1 \cos\phi_B , \: \: \Xi _2 \longrightarrow \Xi _2 \cos\phi_B 
\end{equation}
if the interaction is made with a rotation around the $z$ axis; and the result
of the interaction of qubit $B$ with its control is that in $\rho $
\begin{equation}
\Xi _2 \longrightarrow \Xi _2 \cos\phi_B, \: \: \Xi _3 \longrightarrow \Xi _3 \cos\phi_B 
\end{equation}
if the interaction is made with a rotation
around the $x$ axis. The terms with $\sin\, \phi $ cancel out because there is
an equal mixture of parts with $\phi $ and parts with $-\phi $.

The changes in the state of qubits $A$ and $B$ can be described equivalently by
maps of mean values that describe the state: the result of the interaction of
qubit $A$ with qubit $C$, described by Eq.~(\ref{UD}), is that
\begin{eqnarray}
\langle \Sigma_1 \rangle & \longrightarrow  & \langle \Sigma_1 \rangle \cos\phi_A \nonumber \\
\langle \Sigma_2 \rangle & \longrightarrow  & \langle \Sigma_2 \rangle \cos\phi_A \nonumber \\
\langle \Sigma_1 \Xi_k \rangle &\longrightarrow & \langle \Sigma_1 \Xi_k \rangle \cos\phi_A \nonumber \\
\langle \Sigma_2 \Xi_k \rangle & \longrightarrow  & \langle \Sigma_2 \Xi_k \rangle \cos\phi_A 
\end{eqnarray}
for $k = 1, 2, 3$; the result of the interaction of qubit $B$ with its control is that
\begin{eqnarray}
\langle \Xi _1 \rangle & \longrightarrow  & \langle \Xi _1 \rangle \cos\phi_B \nonumber \\
\langle \Xi _2 \rangle & \longrightarrow  & \langle \Xi _2 \rangle \cos\phi_B \nonumber \\
\langle \Sigma_j \Xi _1 \rangle & \longrightarrow  & \langle \Sigma_j \Xi _1 \rangle \cos\phi_B \nonumber \\
\langle \Sigma_j \Xi _2 \rangle & \longrightarrow  & \langle \Sigma_j \Xi _2 \rangle \cos\phi_B 
\end{eqnarray}
for $j = 1, 2, 3$ if the interaction is made with a rotation around the $z$ axis; and the result of the interaction of qubit $B$ with its control is that
\begin{eqnarray}
\langle \Xi _2 \rangle & \longrightarrow  & \langle \Xi _2 \rangle \cos\phi_B \nonumber \\
\langle \Xi _3 \rangle & \longrightarrow  & \langle \Xi _3 \rangle \cos\phi_B \nonumber \\
\langle \Sigma_j \Xi _2 \rangle & \longrightarrow  & \langle \Sigma_j \Xi _2 \rangle \cos\phi_B \nonumber \\
\langle \Sigma_j \Xi _3 \rangle & \longrightarrow  & \langle \Sigma_j \Xi _3 \rangle \cos\phi_B 
\end{eqnarray} 
for $j = 1, 2, 3$ if the interaction is made with a rotation around the $x$ axis.

When $\phi_A $ and $\phi_B $ change over intervals from initial values
$\phi_{Ai} $ and $\phi_{Bi} $ to final values $\phi_{Af} $ and $\phi_{Bf} $, the
$\cos\phi_A $ and $\cos\phi_B $ factors in the maps are replaced by
$\cos\phi_{Af} /\cos\phi_{Ai} $ and $\cos\phi_{Bf} /\cos\phi_{Bi} $. If either
of these factors is larger than $1$, the map is not completely positive and does
not apply to all density matrices $\rho $ for qubits $A$ and $B$. This happens
whenever the entanglement increases. It also happens in cases where the
concurrence (\ref{concur2}) decreases, when one of $\cos\phi_A $ and $\cos\phi_B
$ increases and the other decreases and there is more decrease than increase.
The completely positive maps that decrease the entanglement have already been
described  \cite{ziman05a}.  

\section{Reconciliation}\label{five}

Entanglement being increased by local interactions may seem surprising from perspectives framed by experience in common situations where it is impossible.
Entanglement is not increased by a completely positive map of the state of
two qubits produced by an interaction on one of them. The interaction will
produce a completely positive map if it is with a control whose state is
initially not correlated with the state of the two qubits, as in
Eq.~(\ref{rabdm}). In our examples, that happens only when the initial value of
$\phi $ is $0$ or a multiple of $\pi $. Otherwise, the state of the control is
correlated with the state of the two qubits as in Eq.~(\ref{fulltran}). When a
subsystem is initially correlated with the rest of a larger system that is being
changed by unitary Hamiltonian dynamics, the map that describes the change of
the state of the subsystem generally is not completely positive and applies to a
limited set of subsystem states  \cite{jordan04a,jordan06a}. We see this in our
examples whenever the entanglement increases and in some cases when the
entanglement decreases.

The map depends on both the dynamics and the initial correlations. It describes
the effect of both in one step. Completely positive maps are what you get in the
simplest set-up where you bring a system and control together in independent
states and consider the effect of the dynamics that begins then. Dynamics over
intervals where the maps are not completely positive can be expected to play
roles in more complex settings. We should not let expectations for completely
positive maps prevent us from seeing things that can happen.

Our perspective is enlarged when we look beyond the map and include the
dynamics. We can see the dynamics and the initial preparation as two related but
separate steps. We can consider the effect of the dynamics, whatever the
preparation may be. 

Local interactions that increase entanglement are completely outside a perspective that is limited to pure states. An interaction on one of the qubits can not change the entanglement at all if
the state of the two qubits remains pure  \cite{bennett96b}. The entanglement of
a pure state of two qubits depends only on the spectrum of the reduced density
matrices that describe the states of the individual qubits, which is the same
for the two qubits. If that could be changed by an interaction on one of the
qubits, there could be a signal faster than light. In our examples, the state of
the two qubits is pure only when it is maximally entangled. In our examples, the
spectrum of the density matrices for the individual qubits never changes, and
gives no measure of the entanglement.

\acknowledgments

We are grateful to a referee for very helpful suggestions, including the comparison with a swap operation. Anil Shaji acknowledges the support of the US Office of Naval Research
through Contract No.~N00014-03-1-0426.

\bibliography{entanglement}
\end{document}